\documentclass[11 pt,oneside,onecolumn,a4paper]{article}
\usepackage{amsmath}
\usepackage{graphicx}
\DeclareGraphicsExtensions{.eps,.ps,.pdf}
\usepackage{bbm}
\usepackage{bm}
\usepackage{epsfig}
\usepackage[T1]{fontenc}
\usepackage{esint}
\usepackage{amssymb}

\usepackage{lipsum}
\usepackage{savesym}

mm \leftmargin 5pt \rightmargin 5pt \hoffset= -15mm

\title{Sequences of information backflow 
in local dephasing channels with spectral gaps}
\author{Filippo Giraldi}

\date{\small{School of Chemistry and Physics, University of KwaZulu-Natal\\ 
and National Institute for Theoretical Physics (NITheP)\\
Westville Campus, Durban 4000, South Africa
\vspace{1em}\\Gruppo Nazionale per la
Fisica Matematica (GNFM-INdAM)\\
c/o Istituto Nazionale di Alta Matematica Francesco Severi\\
Citt\'a Universitaria, Piazza Aldo Moro 5, 00185 Roma, Italy}}

\begin{document}

\maketitle

\def\bbm[#1]{\mbox{\boldmath$#1$}}

\vspace{0em}




\vspace{0em}

\begin{abstract}
The flow of quantum information in local dephasing channels is analyzed over short and long times in case the structured reservoirs of frequency modes exhibit a spectral gap in the density of modes over low frequencies. The presence of the low-frequency gap with upper cut-off frequency $\omega_g$ produces over the time scale $1/\omega_g$ an infinite sequence of time intervals over which information backflow appears. Such time intervals are generally irregular but, under certain conditions, exhibit the following bounds: the $n$th backflow has certainly started at the instant $\pi \left(1+2(n-1)\right)/\omega_g$, and certainly ended at the instant $2\pi n/\omega_g$, for every $n=1,2,\ldots$. The intervals become regular over long times, tend to the asymptotic length $\pi/\omega_g$ as supremum value, and are described analytically in terms of the structure of the spectral density near the cut-off frequency. Consequently, engineering structured reservoirs of frequency modes with low-frequency spectral gaps produces in local dephasing channels regular and controllable sequences of information backflow and recoherence over long times, along with non-Markovian evolution.
\end{abstract}

\maketitle



\section{Introduction}\label{1}

In recent years great deal of attention has been devoted to the interpretation of the non-Markovian dynamics of an open quantum system in terms of the flow of quantum information between the system and the external environment 
\cite{BNnMarkovRMP2016,PNnMarkovRPP2014}. The flow of information has been characterized in the most various ways, via Fisher information \cite{IFlowDef1}, mutual information \cite{IFlowDef2}, or fidelity \cite{IFlowDef3}, to name a few. The trace-distance measure introduced in Ref. \cite{BnnMarkovPRL2009} evaluates the relative distinguishability of two generic quantum states. This measure decreases in Markovian evolution. In this way, the Markovian process can be interpreted as a loss of quantum information in the open system. On the contrary, non-Markovian dynamics can be red as a flow of information from the environment back into the open system.

The flow of quantum information has been largely studied on the referential system of a qubit that experiences pure dephasing and no energy dissipation. Such system is also referred as a local dephasing channel \cite{MPRAr2013,MPRA2014,QbtMPRA2014,MNJP2015,YPRB2003}. For this system the measure of non-Markovianity that is derived from the trace-distance measure \cite{BnnMarkovPRL2009} is given by a simple integral form and the backflow of quantum information is witnessed by persistent negative values of the dephasing rate \cite{MPRAr2013,MPRA2014,QbtMPRA2014,MNJP2015,nnMarkovNeg2LZPRA2011}. The search for ways of generating and controlling the information backflow is a central issue in quantum information processing. A powerful way to control the decoherence process in open quantum systems is the adoption of external controllable interactions, also referred as dynamical decouplings \cite{DDV1,DDV2}. In Ref. \cite{MNJP2015} it is shown how dynamical decoupling pulses reverse and control the information flow and induce non-Markovian evolution. In absence of any external interaction, the information backflow depends strongly on the environmental spectrum and, consequently, can be controlled by engineering the reservoirs. The analysis performed in Refs. \cite{MPRAr2013,MPRA2014,QbtMPRA2014} has shown how the information backflow depends on the features of the environmental spectrum. In fact, the appearance of information backflow is conditioned by convexity properties of the spectral density (SD) of the system \cite{MPRAr2013,MPRA2014,QbtMPRA2014,GPRA2017}. Especially, for SDs that are ohmic-like at low frequency, information backflow appears at zero temperature uniquely if the ohmicity parameter exceeds the value 2. See Refs. \cite{MPRAr2013,MPRA2014,QbtMPRA2014} for details. The low-frequency structure of the SD determines the features of the long-time information flow and regular patters are found in correspondence of periodical intervals of the ohmicity parameter for a large variety of SDs that include and continuously depart from the ohmic-like profiles \cite{GPRA2017}.

Photonic band gap materials play a central role in the realization of anomalous effects in internal dynamics of atoms. As matter of fact, in photonic band gap materials an atomic transition can be coupled to a continuous distribution of frequency modes with spectral gaps  
\cite{PBGYJP1993,PCJbook}. In this way, a variety of phenomena concerning control of atomic decay, population trapping, atomic coherence and atom-photon bound states are obtained, to name a few \cite{PBGJWPRB1991,PBGYJP1993,KJMO1994,PBG1,GPRA1997,PBGLNNBRPP2000,PBGWJJOB2003,MMPSGPRA2009,GSDGPRA2012}. For a qubit that is coupled to a coherent photonic-band-gap reservoir the band gap width influences the quantum speed limit and the memory time of the environment \cite{PBGnnMarkov1}.

In Ref. \cite{PBGnnMarkov2} the influence of SDs with band gaps on the information and energy flow has been analyzed in bosonic finite-chain environments. Such study has shown that the largest backflow of information is obtained at the edges of the gaps and that memory effects become stronger as the ohmicity parameter of algebraic SDs departs from unity. Refer to \cite{PBGnnMarkov2} for details. In local dephasing channels information backflow, recoherence and non-Markovianity occur for persistent negative values of the dephasing rate \cite{MPRAr2013,MPRA2014,QbtMPRA2014}. In these systems the temporal behavior of the information flow is strictly correlated to the environmental spectrum and, for ohmic-like SDs, the ohmicity parameter determines the time intervals at which the information flows back to the open system. 
See Refs. \cite{MPRAr2013,MPRA2014,QbtMPRA2014} for details.

As a continuation of the scenario described above, here we consider structured reservoirs of field modes that exhibit a spectral gap at low frequencies. The SDs under study are variously tailored near the frequency gap, include the inverse-power-law divergences of the periodic dielectric structures and contain perturbations of such feasible profiles. In these conditions we study if and how the spectral gap influence influences the appearance of information backflow in local dephasing channels.

The paper is organized as follows. Section \ref{2} is devoted to the description of local dephasing channels. In Section \ref{3} the SDs with low-frequency gap under study are defined. In Section \ref{4} the short- and long-time behavior of the dephasing rate are analyzed. Section \ref{5} is devoted to the analytical description of the time intervals over which the information flows back into the open system. Summary and conclusion are drawn in Section \ref{6}. Details of the calculations are provided in the Appendix.

\section{Local dephasing channels}\label{2}

Consider a qubit that is coupled to a reservoir of field modes according to the Hamiltonian \cite{MPRAr2013,MPRA2014,QbtMPRA2014}
\begin{equation}
H= \omega_0 \sigma_z+ \sum_k \omega_k b^{\dagger}_k b_k
+\sum_k \sigma_z \left(g_k b_k+g^{\ast}_k b^{\dagger}_k\right), \label{H}
\end{equation}
in units where $\hbar=1$. The parameter $\omega_0$ represents the transition frequency of the qubit, while 
$\sigma_z$ is the $z$-component Pauli spin operator \cite{BP,Weiss}. The index $k$ runs over the frequency modes. The parameter $\omega_k$ is the frequency of the $k$th mode, while $b^{\dagger}_k$ and $b_k$ represent the rising and lowering operator, respectively, of the same mode. The coefficient $g_k$ is the coupling strength between the qubit and the $k$th frequency mode. The reduced density matrix $\rho(t)$ represents the mixed state of the qubit at the time $t$ and is obtained by tracing the density matrix of the whole system at the time $t$ over the Hilbert space of the external environment \cite{BP}. The model is exactly solvable \cite{RH1,RH2,RH3} and describes a pure dephasing process of the qubit.

The initial state of the qubit is decoupled from the reservoir of field modes that is considered to be at zero temperature. In the interaction picture the reduced dynamics is described by the master equation
\begin{equation}
\dot{\rho}(t)=\gamma_0(t)\left(\sigma_z \rho(t) \sigma_z -\rho(t)\right).  \label{Eq1}
\end{equation}
The function $\gamma_0(t)$ represents the dephasing rate and reads
\begin{eqnarray}
&&\gamma_0(t)=\int_{0}^{\infty} \frac{J\left(\omega\right)}{\omega}\,
\sin \left(\omega t \right) \, d\omega.
  \label{gamma0} 
\end{eqnarray}
The function $J\left(\omega\right)$ is the SD of the system and is defined in terms of the coupling constants $g_k$ via the following form,
$J\left(\omega\right)=\sum_k \left|g_k\right|^2 \delta \left(\omega-\omega_k\right)$. The off-diagonal element $\rho_{0,1}(t)$ of the density matrix estimates the quantum coherence between the states $|0\rangle$ and $|1\rangle$ of the qubit and evolves as below \cite{RH1,RH2,RH3},
\begin{equation}
\rho_{0,1}(t)=\rho^{\ast}_{1,0}(t)=\rho_{0,1}(0)\, \exp \left(-\Xi_0(t)\right).  \label{rho01t}
\end{equation}
The function $\Xi_0(t)$ is the dephasing factor and at zero temperature reads 
\begin{eqnarray}
&&\Xi_0(t)=\int_{0}^{\infty} J\left(\omega\right)\, \frac{1- \cos\left(\omega t \right) }{\omega^2}\, d\omega.  \label{Xi0t}
\end{eqnarray}
The full or partial loss of coherence, decoherence or recoherence process are determined by the structured reservoir of field modes and by the couplings via the SD and have been widely studied in literature. See Refs. \cite{MPRAr2013,MPRA2014,QbtMPRA2014,GPRA2017} for details.

For the present system, the appearance of non-Markovian evolution can be studied from the dephasing rate. In fact, the trace-distance measure introduced in Ref. \cite{BnnMarkovPRL2009} provides for the system under study a simple form of the measure of non-Markovianity, \cite{nnMarkovNeg2LZPRA2011,FPRA2013}
\begin{equation}
\mathcal{N}=\int_{\gamma_0(t)<0}\left|\gamma_0(t)\right|e^{-\Xi_0(t)} d t. \label{N}
\end{equation}
Persistent negative values of the dephasing rate contribute to the above measure and correspond to information backflow. The appearance of information backflow in local dephasing channels is determined by the features of the reservoir of field modes and by the couplings. Since the SD determines the temporal behavior of the dephasing rate, great interest has been devoted to the structures of the SD that provide negative values of the dephasing rate. For ohmic-like SDs the information backflow, recoherence and non-Markovian dynamics depend on the ohmicity parameter \cite{MPRAr2013,MPRA2014,QbtMPRA2014,GPRA2017}. We intend to study how the information flow is influenced by a low-frequency gap in the distribution of frequency modes for a large variety of SDs that is defined below.

\section{ Spectral densities with low-frequency gaps}\label{3}

From the theory of photonic band gap materials it is known that periodic dielectric structures provide distribution of frequency modes with gaps. Let $\omega_g$ be the upper cut-off frequency of the band gap. Near the cut-off frequency, $\omega\to\omega_g^{+}$, the density of frequency modes diverges according to forms which are approximately proportional to the expression $
\left(\omega-\omega_g\right)^{\left(1-d_0\right)/2}\Theta\left(\omega-\omega_g\right)$, where $\Theta(\omega)$ is the Heaviside step function. The parameter $d_0$ indicates the dimension of the surface of the Brillouin zone that is spanned by the band edge modes with vanishing group velocity. The constraint $d_0\leq 2$ holds in photonic band gap materials. If the coupling function varies slowly at the edge of the frequency gap, the SD exhibits for $\omega \to \omega_g^+$ the same inverse-power-law divergence as the density of frequency modes does. Refer to \cite{PBGLNNBRPP2000,KJMO1994,PCJbook,PBGJWPRB1991,PBG1} for details.

The structured reservoirs under study consist in continuous distributions of frequency modes with a spectral gap at low frequencies. Let $\omega_g$ be the upper cut-off frequency of the gap. Due to the continuity, the SDs are required to fulfill the constraint \cite{SDPlenio1,ReedSimonBook}
\begin{eqnarray}
\int_{\omega_{\textrm{min}}}^{\omega_{\textrm{max}}} \frac{J\left(\omega\right)}{\omega}\, d\omega<\infty. \label{ConstrSD0}
\end{eqnarray}
The domain of integration $\left[\omega_{\textrm{min}},\omega_{\textrm{max}}\right]$ is the support of the SD. In this case the lower extremum is the upper cut-off frequency of the spectral gap, $\omega_{\textrm{min}}=\omega_g$, and the upper extremum $\omega_{\textrm{max}}$ can be either finite or infinite. The SDs under study vanish below the edge of the low-frequency gap, $J\left(\omega\right)=0$ for $0\leq \omega<\omega_g$. For $\omega\geq \omega_g$ the SDs are characterized by the dimensionless auxiliary function $\Omega\left(\nu\right)$. Such function fulfills the scaling property $J \left( \omega_g+ \omega_s\nu \right)=\omega_s\Omega\left(\nu \right)$ for every $\nu\geq 0$, where $\omega_s$ is a typical scale frequency of the system. We consider two general classes of SDs that are tailored near the cut-off frequency $\omega_g$ according to power laws. The low-frequency behaviors under study include the integrable divergences that characterize the photonic band gap materials. Possible perturbations of such low-frequency profiles are also considered by introducing powers of logarithmic functions as additional factors for the power laws \cite{GPRA2017,GEPJD2015}. Such logarithmic powers are natural valued for the first class, and are arbitrarily real for the second class of SDs under study. In this way, we are equipped to evaluate how stable the information flow is under perturbations of the power-law profiles \cite{GPRA2017}. The two classes of SDs are defined below. 

\subsection{First class of spectral densities}\label{SD1}

The first class of SDs under study is defined by auxiliary functions which are continuous for every $\nu> 0$ and exhibit as $\nu\to 0^+$ the asymptotic behavior \cite{BleisteinBook}
\begin{eqnarray}
&&\hspace{-0em}\Omega\left(\nu\right)\sim 
\sum_{j=0}^{\infty}
\sum_{k=0}^{n_j}c_{j,k} \nu^{\alpha_j}\left(- \ln \nu\right)^k,  \label{o0log} 
\end{eqnarray}
where $\alpha_0>-1$, $\infty> n_j\geq 0$, $\alpha_{j+1}>\alpha_j$ for every $j\geq 0$, and $\alpha_j\uparrow +\infty$ as $j\to +\infty$. The coefficients $c_{j,k}$ must be chosen in such a way that the auxiliary functions $\Omega\left(\nu\right)$ are 
non-negative. Consequently, the constraint $c_{0,n_0}>0$ is 
required. Dominant power-law behavior is obtained for $n_0=0$, and negative values of the power $\alpha_0$, i.e., $0 >\alpha_0>-1$, provide integrable inverse-power-law divergences in $\nu=0$. Such configurations refer to the singularities of the photonic band gap materials at the band gap edge. For $\alpha_0>0$ the logarithmic singularity in $\nu=0$ is removed by defining $\Omega(0)=0$. The support of the SDs can be either finite, $\omega_{\textrm{max}}<\infty$, or infinite, $\omega_{\textrm{max}}=\infty$. In the latter case the asymptotic behavior $\Omega\left(\nu\right)= \mathcal{O}\left(\nu^{-1-\chi_0}\right)$ is required as $\nu\to+\infty$, where $\chi_0>0$. This condition guarantees the integrability of the SDs over high frequencies. The constraint (\ref{ConstrSD0}) is fulfilled. Further necessary conditions involve the function $\Lambda\left(\nu\right)$ that is defined for every $\nu\geq 0$ in terms of the auxiliary function $\Omega\left(\nu\right)$ as
\begin{equation}
\Lambda\left(\nu\right)=\frac{\Omega\left(\nu\right)}{1+\nu/\nu_0}, 
\label{Lambda}
\end{equation}
where $\nu_0=\omega_g/\omega_s$. As matter of facts, the Mellin transforms $\hat{\Lambda}\left(s\right)$ and the meromorphic continuations \cite{BleisteinBook,Wong-BOOK1989} are required to decay sufficiently fast as $\left|\mathrm{Im} \,s \right|\to+\infty$. See Appendix and Ref. \cite{GXiv2016} for details.

The definition of the first class of SDs is quite simple but the logarithmic powers are restricted to natural values. The second class of SDs, which is defined below, includes arbitrarily positive or negative, or vanishing powers of logarithmic forms. Such arbitrariness requires more constraints but allows to perturb the power-law profiles of the SDs near the frequency gap with arbitrarily small, positive or negative, logarithmic powers. In this way, one can evaluate the accuracy of the results that are obtained for the experimentally feasible reservoirs with respect to logarithmic perturbations of the low-frequency power-law profiles \cite{GPRA2017}.

\subsection{Second class of spectral densities}\label{SD2}

The second class of SDs under study is described by auxiliary functions which exhibit as $\nu\to 0^+$ the asymptotic expansion
\begin{eqnarray}
&&\hspace{-0em}\Omega\left(\nu\right)\sim \sum_{j=0}^{\infty}w_j
\, \nu^{\alpha_j} \left(-\ln \nu\right)^{\beta_j}.  \label{OmegaLog0}
\end{eqnarray}
The powers $\alpha_j$ fulfill the constraints that are reported in Sec. \ref{SD1}, while the logarithmic powers $\beta_j$ are arbitrarily real, either positive or negative, or vanishing. Again, dominant power-law behavior is obtained for $\beta_0=0$, and negative values of the power $\alpha_0$, i.e., $0 >\alpha_0>-1$ include the configurations of photonic band gap materials. The corresponding divergences in $\nu=0$ are integrable. Since the auxiliary functions $\Omega\left(\nu\right)$ are non-negative, the coefficients $w_j$ must be chosen accordingly, and the constraint $w_0>0$ is required. Again, for $\alpha_0>0$ the logarithmic singularity in $\nu=0$ is removed by setting $\Omega(0)=0$. The constraint (\ref{ConstrSD0}) is fulfilled. Arbitrarily small, positive (negative) values of the first logarithmic power $\beta_0$ result in arbitrarily small increases (decreases) in the power-law profiles of the edge of the spectral gap. The functions $\Omega\left(\nu\right)$ must be continuous and differentiable in the support and summable. Additional necessary conditions are required for the function $\Lambda\left(\nu\right)$ that is defined by Eq. (\ref{Lambda}). Let $\bar{n}$ be the least natural number such that $\bar{n}\geq\alpha^{\prime}_{\bar{k}}$, where $\alpha^{\prime}_{\bar{k}}$ is the least of the powers $\alpha^{\prime}_k$ that are larger than or equal to unity. The powers $\alpha^{\prime}_k$ are defined in the Appendix in terms of the powers $\alpha_j$. The function $\Lambda^{\left(\bar{n}\right)}\left(\nu\right)$ is the $\bar{n}$th derivative of the function $\Lambda\left(\nu\right)$ and is required to be continuous on the interval $\left(0,\infty\right)$. The integral $\int_0^{\infty}\Lambda\left(\nu\right)\exp\left(-\imath \xi \nu\right) d \nu$ has to converge uniformly for all sufficiently large values of the variable $\xi$ and the integral $\int \Lambda^{\left(\bar{n} \right)}\left(\nu\right)\exp\left(-\imath \xi \nu\right) d \nu$ must converge at $\nu=+\infty$ uniformly for all sufficiently large values of the variable $\xi$. The function $\Lambda\left(\nu\right)$ is required to be differentiable $k$ times and the corresponding derivatives must fulfill as $\nu\to 0^+$ the asymptotic expansion
$$\Lambda^{(k)}\left(\nu\right)\sim \sum_{j=0}^{\infty}w_j
\, \frac{d^k}{d\nu^k}\left(\nu^{\alpha_j} \left(-\ln \nu\right)^{\beta_j}\right),$$
for every $k=0,1, \ldots,\bar{n} $, where $\bar{n}$ is the non-vanishing natural number defined above. Additionally, for every $k=0, \ldots,\bar{n}-1$, the function $\Lambda^{(k)}\left(\nu\right)$ must vanish in the limit $\nu\to +\infty$. The above constraints are based on the asymptotic analysis performed in Ref. \cite{WLJMAA1978}. The SDs under study are non-negative, bounded and summable, due to physical grounds, and, apart from the above constraints, are arbitrarily shaped.

\section{Dephasing rate for structured reservoirs with spectral gaps}\label{4}

We want to investigate the role of the spectral gap in the appearance of the information backflow, recoherence and non-Markovianity in local dephasing channels. Led by such perspective, we analyze the short- and long-time behavior and the sign of the dephasing rate for the two classes of SDs introduced in Section \ref{2}.

If the reservoir of frequency modes exhibits the mentioned low-frequency gap with upper cut-off frequency $\omega_g$, the dephasing rate is given by
\begin{eqnarray}
&&\gamma_0(t)=\int_{\omega_g}^{\infty} \frac{J\left(\omega\right)}{\omega}\,
\sin \left(\omega t \right) \, d\omega.
  \label{gamma0g} 
\end{eqnarray}
The action of the spectral gap can be analyzed by expressing the dephasing rate via the form 
 \begin{eqnarray}
&&\gamma_0(t)=\varphi_c(t) \sin \left(\omega_g t\right)+\varphi_s(t) \cos \left(\omega_g t\right). \label{gammaSC}
\end{eqnarray}
The functions $\varphi_c(t)$ and $\varphi_s(t)$ are defined 
in terms of the SD as 
\begin{eqnarray}
&&\hspace{-1em}\varphi_c(t) =\int_0^{\infty}\frac{J\left(\omega_g+ \omega^{\prime}\right)}{\omega_g+ \omega^{\prime}}\, \cos\left(\omega^{\prime} t\right) d \omega^{\prime}, \label{phc} \\
&&\hspace{-1em}\varphi_s(t) =\int_0^{\infty}\frac{J\left(\omega_g+ \omega^{\prime}\right)}{\omega_g+ \omega^{\prime}}\, \sin\left(\omega^{\prime} t\right) d \omega^{\prime}. \label{phs}
\end{eqnarray}
The expression (\ref{gammaSC}) suggests that two time scales, $1/\omega_s$ and $1/\omega_g$, are involved in the temporal behavior of the dephasing rate. Both the time scales are determined by the SD and depend on the structure of the reservoir of field modes and on the coupling strength. If the following constraint is fulfilled, $\varphi^2_c(t)+\varphi^2_s(t)>0$ for every $t \geq 0$, the expression (\ref{gammaSC}) of the dephasing factor is equivalent to the form
\begin{eqnarray}
&&\gamma_0(t)=\sqrt{\varphi^2_c(t)+\varphi^2_s(t)}
 \sin \left(\omega_g t +\phi(t)\right). \label{gammaS}
\end{eqnarray}
The time-dependent angle $\phi(t)$ vanishes for $t=0$, i.e., $\phi(0)=0$, and is defined for $t>0$ via the functions $\varphi_c(t)$ and $\varphi_s(t)$ as below,
\begin{eqnarray}
&&\hspace{-1em}\phi(t)=0,\hspace{1em}\text{if}\hspace{1em} \varphi_s(t)=0 \hspace{1em}\text{and}\hspace{1em}\varphi_c(t)>0,\label{phi0} \\
&&\hspace{-1em}\phi(t)=\pi,\hspace{1em}\text{if}\hspace{1em} \varphi_s(t)=0 \hspace{1em}\text{and}\hspace{1em}\varphi_c(t)<0,\label{phi1} \\
&&\hspace{-1em}\phi(t)=\operatorname{arccot} \frac{\varphi_c(t)}{\varphi_s(t)},\hspace{1em}\text{if}\hspace{1em} \varphi_s(t)>0,\label{phi2} \\
&&\hspace{-1em}\phi(t) =\pi +\operatorname{arccot} \frac{\varphi_c(t)}{\varphi_s(t)},\hspace{1em}\text{if}\hspace{1em} \varphi_s(t)<0. \label{phi3}
\end{eqnarray}
In general, the angle $\phi(t)$ is not a continuous function of time. On the contrary, the function $\phi(t)$ is continuous for every $t\geq 0$ and differentiable for $t>0$ if the sine transform $\varphi_s(t)$ is positive, $\varphi_s(t)>0$ for every $t > 0$, and if the cosine transform $\varphi_c(t)$ and the sine transforms $\varphi_s(t)$ are continuous for every $t \geq0$, and 
differentiable for every $t>0$. 

Consider regular SDs such that the corresponding functions $\varphi_c(t)$ and $\varphi_s(t)$ are continuous and differentiable, and such that the constraint
\begin{eqnarray}
J^{\prime}\left(\omega\right)<\frac{J\left(\omega\right)}{\omega}. \label{conSDphisp}
\end{eqnarray}
holds for every $\omega>\omega_g$. For such SDs the sine transform $\varphi_s(t)$ is positive, $\varphi_s(t)>0$ for every $t > 0$, and the angle $\phi(t)$, given by Eq. (\ref{phi2}), is a continuous and differentiable function of time. For example, the constraint (\ref{conSDphisp}) is fulfilled by the following SDs that behave as inverse power laws or Lorentzians in the frequency gap,
\begin{eqnarray}
&&\hspace{-2em}J_1\left(\omega_g+ \omega^{\prime}\right) = q_1 \Theta \left(\omega^{\prime}\right) \left(\omega^{\prime}\right)^{\alpha} e^{-\lambda_1 \omega^{\prime}}, \label{j1} \\
&&\hspace{-2em}J_2\left(\omega_g+ \omega^{\prime}\right) =  \frac{q_2 \Theta \left(\omega^{\prime}\right)}{\left(\lambda_2\right)^2+\left(\omega^{\prime}\right)^2}, \label{j2}
\end{eqnarray}
where $0>\alpha>-1$, $q_1>0$, $\lambda_1>0$, $q_2>0$,
 $\lambda_2>0$. The corresponding sine transform $\varphi_s(t)$ is positive for every $t > 0$. Notice that the negative values of the power $\alpha$, appearing in Eq. (\ref{j1}), reproduce the inverse-power-law divergences of the photonic band gap structures in the upper cut-off frequency $\omega_g$.

\subsection{Short- and long-time behaviors}\label{41}

The short-time behavior of the dephasing rate is determined by the high-frequency structure and integral properties of the SD. A common short-time behavior is obtained if the SDs decay sufficiently fast at high frequencies. In fact, 
for $t \ll \min\left\{1/ \omega_s,1/\omega_g\right\}$, the dephasing rate increases linearly,
\begin{equation}
\gamma_0(t) \sim l_0 t,\label{gamma0short0}
\end{equation}
if the SDs belong to the first class and $\chi_0>1$, or to the second class and $\chi_0>3$. The coefficient $l_0$ depends on integral properties of the SD and is defined in the Appendix. The short-time positivity of the dephasing rate suggests that information and coherence are initially lost by the open system and that the short-time dynamics is Markovian. Such behavior appears also in absence of any spectral gap \cite{GPRA2017}. Consequently, we conclude that the spectral gap does not hinder the short-time loss of information or coherence in local dephasing channels.

Over the long times $t \gg 1/\omega_s$ the dephasing rate exhibits damped oscillations. The long-time damping is described by Eq. (\ref{gammaSC}) via the asymptotic forms of the functions $\varphi_c(t)$ and $\varphi_s(t)$, which depend on the low-frequency structure of the SDs. Such asymptotic expansions are analyzed below for the first and second class of SDs that are defined in Secs. \ref{SD1} and \ref{SD2}, respectively. Consider the first class of SDs. If the power $\alpha_0$ is negative, $0>\alpha_0>-1$, or if it is positive, $\alpha_0>0$, and differs from even natural numbers, the function $\varphi_c(t)$ vanishes for $t \gg 1/ \omega_s$ as
\begin{eqnarray}
\varphi_c(t) \sim c_{0,n_0} \mu_c \left(\omega_s t\right)^{-1-\alpha_0}
\ln^{n_0}\left(\omega_s t\right), \label{varphic1}
\end{eqnarray}
where $\mu_c=\omega^2_s \omega^{-1}_g  \cos \left(\pi \left(1+\alpha_0\right)/2\right)\Gamma\left(1+\alpha_0\right)$. The above relaxations reduce to dominant inverse power laws for $n_0=0$,
\begin{eqnarray}
\varphi_c(t) \sim c_{0,0} \mu_c \left(\omega_s t\right)^{-1-\alpha_0}. \label{varphic1p}
\end{eqnarray}
If the power $\alpha_0$ is a non-vanishing even natural number, $\alpha_0=2 m_1$, where $m_1$ is a positive integer, and the logarithmic power $n_0$ does not vanish, the function $\varphi_c(t)$ relaxes for $t \gg 1/\omega_s$ as below,
\begin{eqnarray}
\varphi_c(t)\sim  c_{0,n_0} \mu^{\prime}_c \left(\omega_s t\right)^{-1-2 m_1} \ln^{n_0-1}\left(\omega_s t\right). \label{varphic2}
\end{eqnarray}
The parameter $\mu^{\prime }_c$ reads $\mu^{\prime }_c=\pi \omega^2_s \omega^{-1}_g n_0(-1)^{m_1}\left(2 m_1\right)!\, /2$. The above asymptotic relaxations provide dominant inverse power laws for $n_0=1$,
\begin{eqnarray}
\varphi_c(t)\sim  c_{0,1} \mu^{\prime }_c \left(\omega_s t\right)^{-1-2 m_1}. \label{varphic2p}
\end{eqnarray}
If $\alpha_0=n_0=0$, or if $\alpha_0= 2m_1$ and $n_0=0$, the relaxations are faster and depend on the powers $\alpha^{\prime}_r$ and $n^{\prime}_r$ that are defined in the Appendix in terms of the powers $\alpha_j$ and $n_j$. Consider the least non-vanishing index $k_1$ such that either $\alpha^{\prime}_{k_1}$ is not an even natural number or $\alpha^{\prime}_{k_1}= 2m_{k_1}$, where $m_{k_1}$ and $n^{\prime}_{k_1}$ are non-vanishing natural numbers. We focus on SDs such that the index $k_1$ exists with the required properties. The function $\varphi_c(t)$ is obtained, in the former case, from Eqs. (\ref{varphic1}) and (\ref{varphic1p}) by substituting the power $\alpha_0$ with $\alpha^{\prime}_{k_1}$ and $n_0$ with $n^{\prime}_{k_1}$, and, in the latter case, from Eqs. (\ref{varphic2}) and (\ref{varphic2p}) by substituting the parameter $m_1$ with $m^{\prime}_{k_1}$ and $n_0$ with $n^{\prime}_{k_1}$.

If the SDs belong to the second class under study, we obtain for $t \gg 1/\omega_s$ a variety of relaxations of the function $\varphi_c(t)$ which are similar to those found for the first class. In addition, the relaxations involve arbitrarily positive or negative, or vanishing powers of logarithmic forms,
\begin{eqnarray}
\varphi_c(t)\sim w_0\left(\omega_s t\right)^{-1-\alpha_0} 
\left(\mu_c  \ln^{\beta_0} \left(\omega_s t\right) + \bar{\mu}_c  \ln^{\beta_0-1} \left(\omega_s t\right)\right), \label{C1tInf0SD2}
\end{eqnarray}
where $\bar{\mu}_c=\omega^2_s \omega^{-1}_g\beta_0\left(\pi \Gamma\left(1+\alpha_0\right)\cos\left(\pi \alpha_0/2\right)/2+ \sin\left(\pi \alpha_0/2\right)\Gamma^{(1)}\left(1+\alpha_0\right)\right)$. If the power $\alpha_0$ is negative, $0>\alpha_0>-1$, or if it is positive, $\alpha_0>0$, and differs from even natural values, the dominant part of the above asymptotic form is 
\begin{eqnarray}
\varphi_c(t)\sim w_0 \mu_c \left(\omega_s t\right)
^{-1-\alpha_0} \ln^{\beta_0}\left(\omega_s t\right),
 \label{C1tInf0SD22}
\end{eqnarray}
 and for $\beta_0=0$ gives dominant inverse power laws,
\begin{eqnarray}
\varphi_c(t)\sim w_0 \mu_c \left(\omega_s t\right)
^{-1-\alpha_0}\left(\omega_s t\right).
 \label{C1tInf0SD23}
\end{eqnarray}
If the power $\alpha_0$ is an even natural number and $\beta_0$ does not vanish, Eq. (\ref{C1tInf0SD2}) provides faster logarithmic relaxations,
\begin{eqnarray}
\varphi_c(t)\sim w_0 \bar{\mu}_c\left(\omega_s t\right)
^{-1-\alpha_0} \ln^{\beta_0-1}\left(\omega_s t\right),
 \label{C1tInf0SD24}
\end{eqnarray}
 that result for $\beta_0=1$ in the dominant inverse power laws 
\begin{eqnarray}
\varphi_c(t)\sim w_0 \bar{\mu}_c \left(\omega_s t\right)^{-\alpha_0}.
 \label{C1tInf0SD25}
\end{eqnarray}
If the power $\alpha_0$ is an even natural number and $\beta_0=0$ the relaxations are even faster and are obtained with the method developed above for first class of SDs.

The long-time behavior of the function $\varphi_s(t)$ depends on the structure of the SD near the band-gap edge similarly as the function 
$\varphi_c(t)$ does. Consider the first class of SDs. If $0>\alpha_0>-1$, if $\alpha_0=n_0=0$, or if $\alpha_0$ is positive 
and is not an odd natural number, the function $\varphi_s(t)$ vanishes for $t\gg1/\omega_s$ according to the relaxations as below,
\begin{eqnarray}
\hspace{-0.9em}\varphi_s(t)\sim c_{0,n_0} \mu_s \,\left(\omega_s t\right)^{-1-\alpha_0}
\ln^{n_0}\left(\omega_s t\right), 
 \label{varphis1}
\end{eqnarray}
which become inverse power laws for $n_0=0$,
\begin{eqnarray}
\hspace{-0.9em}\varphi_s(t)\sim c_{0,0} \mu_s \,\left(\omega_s t\right)^{-1-\alpha_0}, 
 \label{varphis1p}
\end{eqnarray}
where $\mu_s=\omega^2_s \omega^{-1}_g\sin\left(\pi \left(1+\alpha_0\right)/2\right)\Gamma\left(1+\alpha_0\right)$. If the power $\alpha_0$ is odd, $\alpha_0=1+2 m_2$, where $m_2$ is a natural number, and $n_0$ is a non-vanishing natural number, the function $\varphi_s(t)$ relaxes for $t \gg 1/\omega_s$ as
\begin{eqnarray}
\varphi_s(t)\sim c_{0,n_0}\mu_s^{\prime} \left(\omega_s t\right)
^{-2-2m_2}
\ln^{n_0-1}\left(\omega_s t\right), \label{varphis2}
\end{eqnarray}
where $\mu_s^{\prime}=(-1)^{m_2}\pi n_0 \left(1+2 m_2\right)!\omega^2_s \omega^{-1}_g /2$. 
The above relaxations turn into dominant inverse power laws if $n_0=1$,
\begin{eqnarray}
\varphi_s(t)\sim c_{0,1}\mu_s^{\prime} \left(\omega_s t\right)
^{-2-2m_2}. \label{varphis2p}
\end{eqnarray}
Faster relaxations appear if the power $\alpha_0$ takes odd natural values and $n_0$ vanishes. Let $k_2$ be the least 
non-vanishing index such that the power $\alpha^{\prime}_{k_2}$ is not an odd natural 
number, or $\alpha^{\prime}_{k_2}=1+ 2m_{k_2}$, where $m_{k_2}$ is natural
 valued, and $n^{\prime}_{k_2}$ is a non-vanishing natural number. We consider
 SDs such that the index $k_2$ exists with the required properties. 
The long-time behavior of the function $\varphi_s(t)$ is obtained, in the former 
case, from Eqs. (\ref{varphis1}) and (\ref{varphis1p}) by substituting 
the parameter $\alpha_0$ with $\alpha^{\prime}_{k_2}$ and $n_0$ with $n^{\prime}_{k_2}$, and, in the latter case, from Eqs. (\ref{varphis2}) and (\ref{varphis2p}) 
by substituting the parameter $m_2$ with $m_{k_2}$ and $n_0$ with $n^{\prime}_{k_2}$.

For the second class of SDs we obtain for $t \gg 1/\omega_s$ a variety of relaxations that are similar to those obtained for the first class but, additionally, involve arbitrarily positive or negative, or vanishing powers of logarithmic forms,
\begin{eqnarray}
\varphi_s(t)\sim w_0 \left(\omega_s t\right)^{-1-\alpha_0}
\left(\mu_s  \ln^{\beta_0}\left(\omega_s t\right)+
\bar{\mu}_s  \ln^{\beta_0-1}\left(\omega_s t\right)
\right), \label{C2tInf0SD2}
\end{eqnarray}
where $\bar{\mu}_s=\beta_0 \omega^2_s \omega^{-1}_g\left(
\pi \sin\left(\pi \alpha_0/2\right) 
\Gamma\left(1+\alpha_0\right)/2-\cos\left(\pi \alpha_0/2\right)
\Gamma^{(1)}\left(1+\alpha_0\right)
 \right)$. If the power $\alpha_0$ takes negative values, $0>\alpha_0>-1$, or if $\alpha_0=\beta_0=0$, or if it is positive 
and is not an odd natural number, the dominant part of the above asymptotic form is 
\begin{eqnarray}
\varphi_s(t)\sim w_0 \mu_s \left(\omega_s t\right)
^{-1-\alpha_0} \ln^{\beta_0}\left(\omega_s t\right),
\label{C2tInf0SD3}
\end{eqnarray}
 and provides for $\beta_0=0$ the inverse power laws 
\begin{eqnarray}
\varphi_s(t)\sim w_0 \mu_s
 \left(\omega_s t\right)^{-1-\alpha_0}.
\label{C2tInf0SD4}
\end{eqnarray}
If the power $\alpha_0$ is an odd natural number and $\beta_0$ does not vanish, Eq. (\ref{C2tInf0SD2}) gives faster logarithmic relaxations,
\begin{eqnarray}
\varphi_s(t)\sim w_0 \bar{\mu}_s\left(\omega_s t\right)
^{-1-\alpha_0} \ln^{\beta_0-1}\left(\omega_s t\right),
\label{C2tInf0SD5}
\end{eqnarray}
 which turn for $\beta_0=1$ into dominant inverse power laws,
\begin{eqnarray}
\varphi_s(t)\sim w_0 
\bar{\mu}_s\left(\omega_s t\right)^{-1-\alpha_0}.
\label{C2tInf0SD6}
\end{eqnarray}
If the power $\alpha_0$ is an odd natural number and $\beta_0=0$ the relaxations are even faster and are obtained with the method developed above for first class of SDs. The analysis of the long-time behavior of the functions $\varphi_c(t)$ and $\varphi_s(t)$ helps to study the sign of the dephasing rate over long times. In this way, the influence of the spectral gap on the information backflow can be studied. %

\section{Sequences of information backflow}\label{5}

The measure of non-Markovianity given by Eq. (\ref{N}) suggests that information backflow appears for persistent negative values of the dehasing rate. The expression (\ref{gammaS}) of the dephasing rate indicates that an infinite sequence of information backflow is obtained if the angle $\phi(t)$ is a continuous function of time, and bounded for $t \to \infty$. We remind that 
if the cosine transform $\varphi_c(t)$ and sine transform $\varphi_s(t)$ are continuous for every $t\geq 0$, and differentiable for every $t>0$, and if the function $\varphi_s(t)$ is positive for every $t>0$, the function $\phi(t)$ is continuous and differentiable and belongs to the interval $\left.\right] 0,\pi\left[\right.$ for every $t>0$. Particularly, the constraint (\ref{conSDphisp}), involving the SD, guarantees the positivity of the sine transform, $\varphi_s(t)>0$ for every $t> 0$.

Temporal bounds can be determined for the intervals over which information flows back in the open system. In fact, under the conditions described in the previous paragraph, the information backflow has already started at the instant $\pi/\omega_g$, 
and the first interval has certainly ended at the instant $2\pi/\omega_g$. Information backflow appears over an infinite sequence of time intervals and, for every $n=1,2,\ldots$, the $n$th backflow has already started at the instant $\pi \left(1+2(n-1)\right)/\omega_g$ and has certainly ended at the instant 
$2\pi n/\omega_g$. Every time interval is no longer than $\pi/\omega_g$. Consequently, the spectral gap with cut-off frequency $\omega_g$ certainly induces non-Markovianity, information backflow and recoherence after the time interval $\pi/\omega_g$, 
and an infinite sequence of time intervals over which the information flows back into the open system.

Over sufficiently long times, the sequence of time intervals over which information backflow appears, becomes regular. An analytical description of such regular sequence 
is provided below in terms of the cut-off frequency $\omega_g$ and of the angle $\phi\left(\infty\right)$. This quantity is defined as the limit of the time-dependent angle $\phi(t)$ for $t \to \infty$. Such limit exists finite since the sign of the sine transform $\varphi_s(t)$ is constant over long times, $\varphi_s(t)>_{\left(<\right)}0$ for $t \gg 1/\omega_s$. Refer to Sec. \ref{41} for details. Consequently, the function $\phi(t)$ is continuous over long times and the limit $\phi\left(\infty\right)$ exists finite and depends on the structure of the SD near the cut-off frequency $\omega_g$.

 Consider the SDs that are defined in Secs. \ref{SD1} and \ref{SD2}. For the the allowed negative values of the power $\alpha_0$, the limit $\phi\left(\infty\right)$ is given by the form
\begin{eqnarray}
&&\hspace{-2em}\phi\left(\infty\right)=
\frac{\pi}{2} \left(1+\alpha_0\right), \hspace{1em}\textrm{if} \hspace{1em} 
0>\alpha_0>-1. \label{phiinf0} 
\end{eqnarray}
 If the power $\alpha_0$ is positive and is not a natural number
 we find
\begin{eqnarray}
&&\hspace{-2em}\phi\left(\infty\right)=\pi\left(\frac{1+\alpha_0}{2}
-\lfloor\frac{1+\alpha_0}{2} \rfloor+\Theta\left(-\cos
\left(\frac{\pi}{2} \alpha_0\right)\right)\right),  \label{phiinf0} \\&&\hspace{-2em} 
\textrm{if} \hspace{1em}\alpha_0>0, \hspace{1em}\textrm{and} \hspace{1em}\alpha_0 \neq \lfloor \alpha_0\rfloor. \nonumber
\end{eqnarray}
If the power $\alpha_0$ takes even natural values the limit $\phi\left(\infty\right)$ is
\begin{eqnarray}
&&\hspace{-2em}\phi\left(\infty\right)=\frac{\pi}{2} \left(2-(-1)^{m}\right),
\hspace{1em}\textrm{if} \hspace{1em}\alpha_0=2 m,\hspace{1em}\textrm{and} \hspace{1em}m=0,1,2,\ldots. \label{phiinf2}
\end{eqnarray}
If the power $\alpha_0$ takes odd natural values and the logarithmic power $n_0$, for the first class of SDs, or $\beta_0$, for the second class of SDs, does not vanish, the limit $\phi\left(\infty\right)$ is given by
\begin{eqnarray}
&&\hspace{-2em}\phi\left(\infty\right)=\frac{\pi}{2} \left(3-(-1)^{m}\right),
\hspace{1em}\textrm{if} \hspace{1em}\alpha_0=1+2 m,\hspace{1em} n_0>0,\label{phiinf3}\\
&&\hspace{-2em}\beta_0\neq 0,\hspace{1em}\textrm{and} \hspace{1em}m=0,1,2,\ldots. \nonumber
\end{eqnarray}
If the power $\alpha_0$ takes odd natural values, $\alpha_0=1+2 m$ with $m=0,1,2,\ldots$, and the logarithmic power $n_0$, for the first class of SDs, or $\beta_0$, for the second class of SDs, vanishes, the limit $\phi\left(\infty\right)$ is determined by the power $\alpha_{k_2}^{\prime}$ that is defined in Sec. \ref{41}. If the power $\alpha_{k_2}^{\prime}$ is not an odd natural number we find
\begin{eqnarray}
&&\hspace{-2em}\phi\left(\infty\right)=
\frac{\pi}{2} \left(2+ \left((-1)^m-1\right)\operatorname{sign}\left(\cos\left(\frac{\pi}{2}\alpha_{k_2}\right)\right)\right),
\label{phiinf4}\\
&&\hspace{-2em}\textrm{if} \hspace{1em}
\alpha_0=1+2 m, \hspace{1em}\textrm{and} \hspace{1em} \,n_0=\beta_0= 0, \nonumber
\end{eqnarray}
while, if $\alpha_{k_2}^{\prime}=1+2 m_{k_2}$ where $m_{k_2}$ is a natural number, the limit is 
\begin{eqnarray}
&&\hspace{-2em}\phi\left(\infty\right)=\frac{\pi}{2} \left(2
+(-1)^{m_{k_2}}\left((-1)^{m}-1\right)\right),\label{phiinf5}
\\&&\hspace{-2em}\textrm{if} \hspace{1em}\alpha_0=1+2 m,
\hspace{1em}\textrm{and} \hspace{1em}n_0=\beta_0= 0. \nonumber
\end{eqnarray}
The above analysis of the limit $\phi\left(\infty\right)$ helps to study the sign of dephasing rate over long times via Eq. (\ref{gammaS}).

\subsection{Regular long-time 
intervals for information backflow}\label{51}

At this stage we are equipped to describe analytically the regular sequence of the long-time intervals over which the information flows back into the open system. Since the limit $\phi\left(\infty\right)$ exists finite, the quantity $\left|\phi(t)-\phi\left(\infty\right)\right|$ vanishes for $t\gg 1/\omega_s$. We focus on long times $t$ fulfilling the constraint 
\begin{eqnarray}
\left|\phi(t)-\phi\left(\infty\right)\right|<\epsilon_0, \label{Cepsilon0}
\end{eqnarray}
where $0<\epsilon_0<\pi/2$. Such long times certainly exist due to the existence of the limit $\phi\left(\infty\right)$. At these times the dephasing rate is certainly negative and information backflow appears over the sequence of intervals $\left[t_{1,n},t_{2,n}\right]$, where
\begin{eqnarray}
&&t_{1,n}=\frac{1}{\omega_g}\left(\pi \left(1+2n\right)- \phi\left(\infty\right)+\epsilon_0\right), \label{t1n} \\
&&t_{2,n}=\frac{1}{\omega_g}\left(2\pi \left(1+n\right)- \phi\left(\infty\right)-\epsilon_0\right). \label{t2n} 
\end{eqnarray}
The natural values $n$ are chosen in such a way that the instants $t_{1,n}$ belong to the long times that are defined via the constraint (\ref{Cepsilon0}). For example, the choice $n\gg\bar{n}$, where $\bar{n}=2+ \lfloor\omega_g/\omega_s\rfloor$, guarantees that the constraint
$t_{1,n}\gg 1/\omega_s$ holds for every $n\gg \bar{n}$, and that the condition $t\gg 1/\omega_s$ is fulfilled for every instant $t$ of the intervals. Notice that the long-time intervals $\left[t_{1,n},t_{2,n}\right]$ are entirely determined by the low-frequency structure of the SD, the cut-off frequency $\omega_g$ and the scale frequency $\omega_s$. The condition $\epsilon_0<\pi/2$ guarantees the existence of each interval, i.e., $t_{1,n}<t_{2,n}$ for every value of $n$.

The cut-off frequency $\omega_g$ and the value of the parameter $\epsilon_0$ determine the length of the long-time intervals over which information flows back into the open system,
\begin{eqnarray}
t_{2,n}-t_{1,n}=\frac{\pi-2 \epsilon_0}{\omega_g}, \label{dt2n}
\end{eqnarray}
for every $n\gg \bar{n}$. As $t \to \infty$ the value $\epsilon_0$ can be chosen as an arbitrarily small, positive number and the duration of each interval tends to the supremum $\pi/\omega_g$ as asymptotic value. Consequently, the intervals over which information backflow appears, tend over very long times to the regular forms 
$$\left.\Bigg]\frac{1}{\omega_g}\left(\pi \left(1+2 n\right)- \phi\left(\infty\right)\right),\frac{1}{\omega_g}\left(2\pi \left(1+ n\right)- \phi\left(\infty\right)\right)\Bigg[\right..$$ The natural values $n$ are chosen several orders of magnitude larger than the value $\bar{n}$ in order to realize the limit $\phi\left(\infty\right)$ for $t \to \infty$. Over such long times the information flows alternatively forth, in the external environment, and back, into the open system, at regular time intervals. The length of the intervals is arbitrarily close to the value $\pi/\omega_g$. Notice that over such times the magnitude of the dephasing rate tends to vanish. We remind that the regular behavior of the information backflow holds for the large variety of SDs with spectral gaps at low frequencies that has been defined in Sec. \ref{3}. Such variety includes the low-frequency inverse-power-law divergences of photonic band gap materials and perturbations of the low-frequency power-law profiles that are described by arbitrary powers of logarithmic forms. Such arbitrariness suggests that the length of the long-time intervals is stable with respect to logarithmic perturbations of the low-frequency power-law profiles and even with respect to changes in the power-laws of the SDs \cite{GPRA2017,GXiv2016}.

\begin{figure}[t]
\centering
\includegraphics[height=6.75 cm, width=9.75 cm]{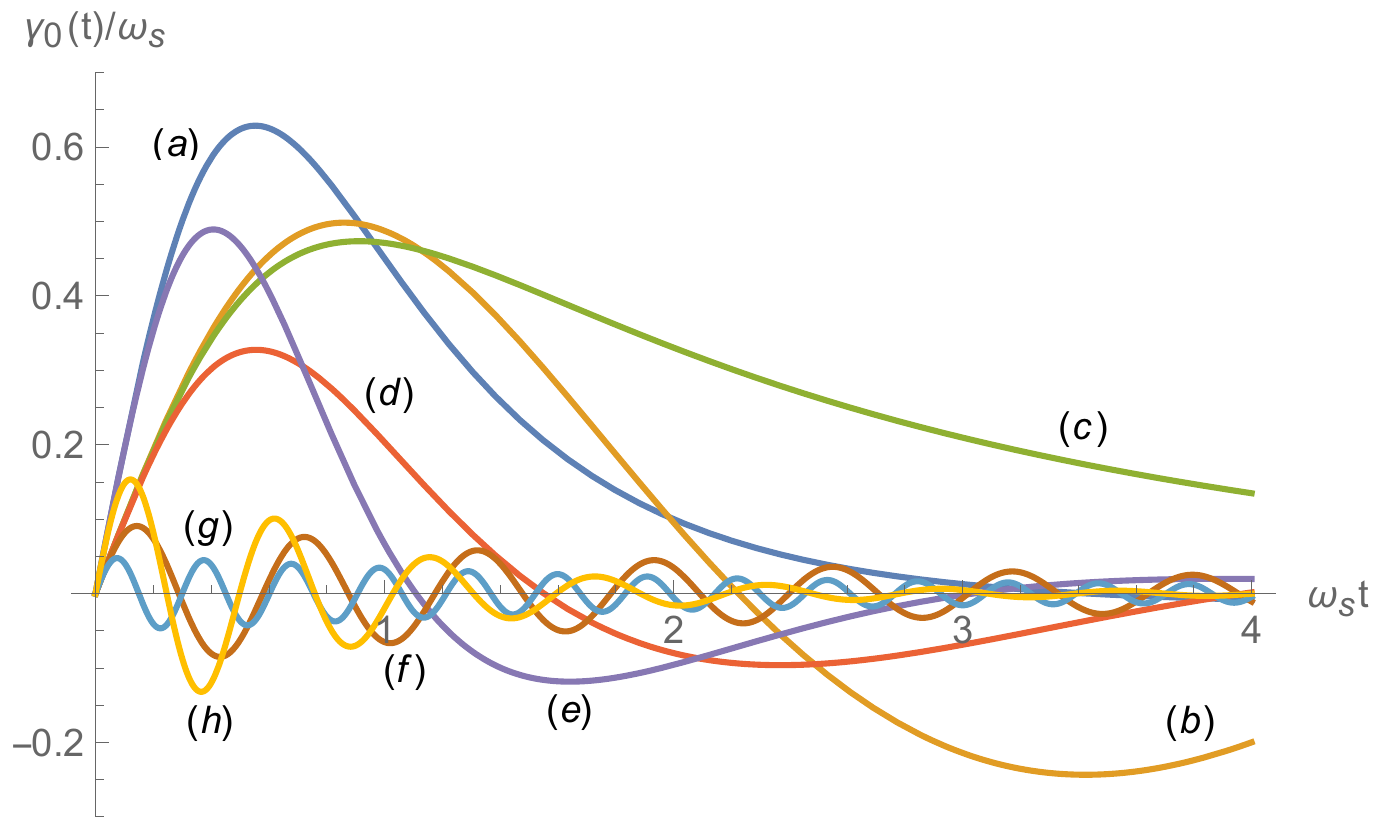}
\vspace*{0cm}
\caption{(Color online) Quantity 
$ \gamma(t)/\omega_s$ versus $\omega_s t$ 
for $0\leq \omega_s t \leq 4$, $J\left(\omega\right)=\omega_s
\left(\omega/\omega_s\right)^{\alpha}\exp\left(- \omega/\omega_s\right)$, 
different values of the ratio $\omega_g /\omega_s$ and of the 
parameter $\alpha$. Curve $(a)$
 corresponds to $\omega_g/\omega_s=0.1$ and $\alpha=2$; 
$(b)$ corresponds to $\omega_g/\omega_s=1$ and $\alpha=0$; 
$(c)$ corresponds to $\omega_g/\omega_s=0.1$ and $\alpha=1$; 
$(d)$ corresponds to $\omega_g/\omega_s=1$ and $\alpha=1$; 
$(e)$ corresponds to $\omega_g/\omega_s=1$ and $\alpha=2$; 
$(f)$ corresponds to $\omega_g/\omega_s=10$ and $\alpha=0$; 
$(g)$ corresponds to $\omega_g/\omega_s=20$ and $\alpha=0$; 
$(h)$ corresponds to $\omega_g/\omega_s=10$ and $\alpha=2$.
}
\label{Fig1}
\end{figure}

\begin{figure}[t]
\centering
\includegraphics[height=6.75 cm, width=9.75 cm]{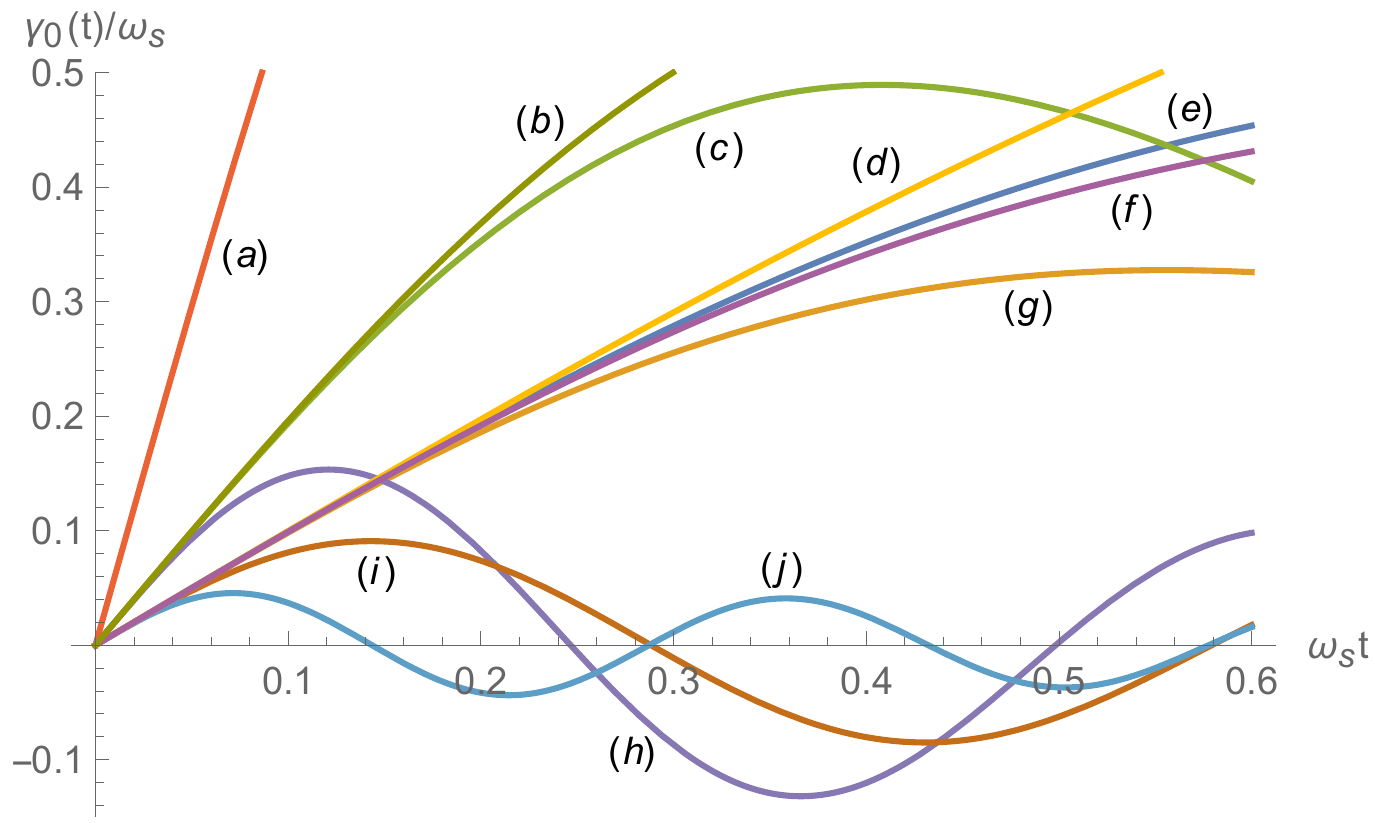}
\vspace*{0cm}
\caption{(Color online) Quantity 
$ \gamma(t)/\omega_s$ versus $\omega_s t$ 
for $0\leq \omega_s t \leq 0.6$, $J\left(\omega\right)=\omega_s
\left(\omega/\omega_s\right)^{\alpha}\exp\left(- \omega/\omega_s\right)$, 
different values of the ratio $\omega_g /\omega_s$ and of the 
parameter $\alpha$. Curve $(a)$
 corresponds to $\omega_g/\omega_s=1$ and $\alpha=3$; 
$(b)$ corresponds to $\omega_g/\omega_s=0.1$ and $\alpha=2$; 
$(c)$ corresponds to $\omega_g/\omega_s=1$ and $\alpha=2$; 
$(d)$ corresponds to $\omega_g/\omega_s=0.1$ and $\alpha=0$; 
$(e)$ corresponds to $\omega_g/\omega_s=1$ and $\alpha=0$; 
$(f)$ corresponds to $\omega_g/\omega_s=0.1$ and $\alpha=1$; 
$(g)$ corresponds to $\omega_g/\omega_s=1$ and $\alpha=1$; 
$(h)$ corresponds to $\omega_g/\omega_s=10$ and $\alpha=2$,
$(i)$ corresponds to $\omega_g/\omega_s=10$ and $\alpha=0$,
$(j)$ corresponds to $\omega_g/\omega_s=20$ and $\alpha=1$.
}
\label{Fig2}
\end{figure}

\begin{figure}[t]
\centering
\includegraphics[height=6.75 cm, width=9.75 cm]{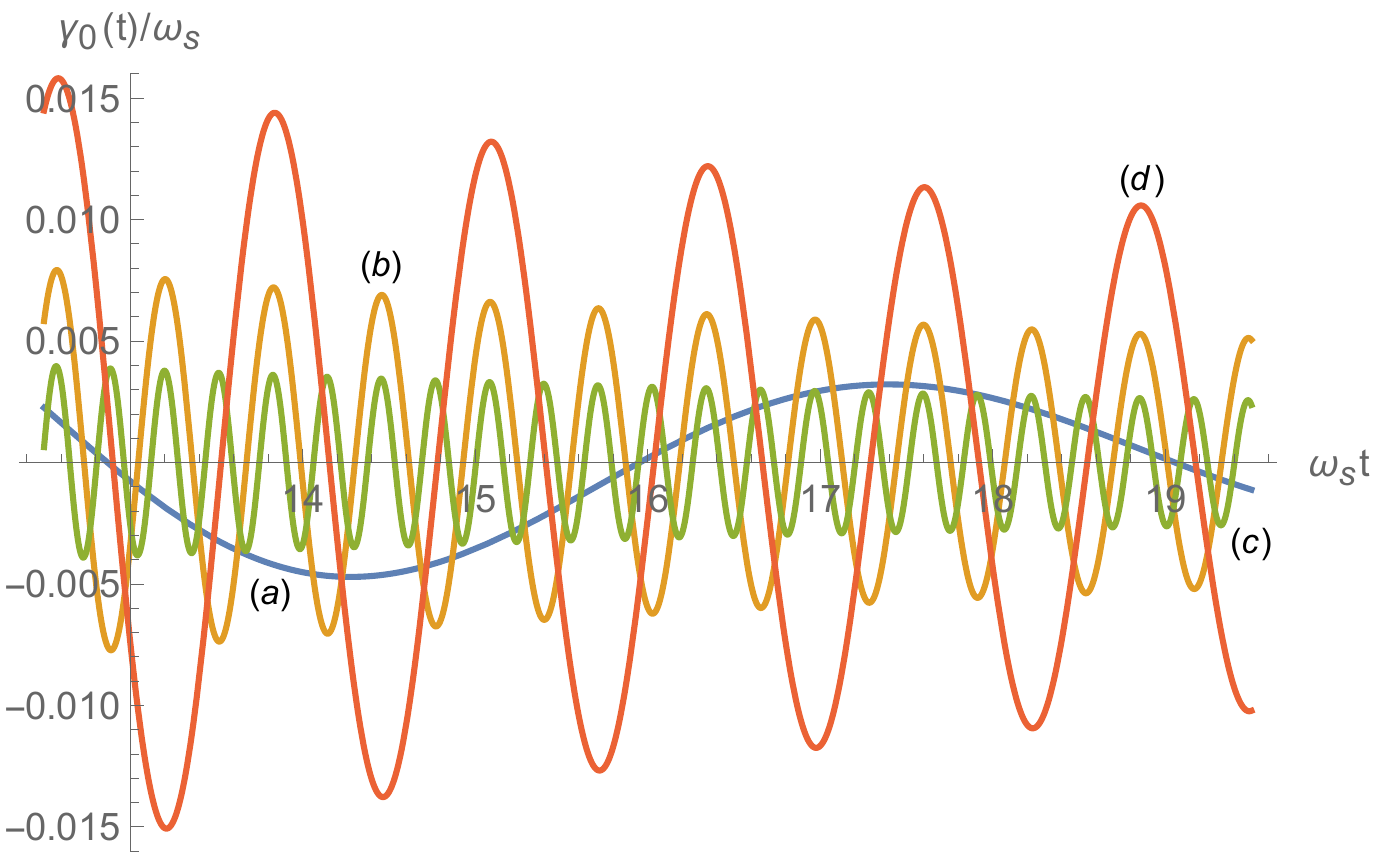}
\vspace*{0cm}
\caption{(Color online) Quantity 
$ \gamma(t)/\omega_s$ versus $\omega_s t$ 
for $12.5\leq \omega_s t \leq 19.5$, $J\left(\omega\right)=\omega_s
\left(\omega/\omega_s\right)^{\alpha}\exp\left(- \omega/\omega_s\right)$, 
different values of the ratio $\omega_g /\omega_s$ and of the 
parameter $\alpha$. Curve $(a)$
 corresponds to $\omega_g/\omega_s=1$ and $\alpha=1$; 
$(b)$ corresponds to $\omega_g/\omega_s=10$ and $\alpha=0$; 
$(c)$ corresponds to $\omega_g/\omega_s=20$ and $\alpha=0$; 
$(d)$ corresponds to $\omega_g/\omega_s=5$ and $\alpha=0$.
}
\label{Fig3}
\end{figure}

Numerical analysis of the dephasing rate is displayed in Figs. \ref{Fig1}, \ref{Fig2} and \ref{Fig3}. The short-time linear growth is shown in Fig. \ref{Fig2}. The long-time oscillations of the depashing rate and the intervals over which information backflow appears (negative values) are displayed in Fig. \ref{Fig3}. The numerical computations are in accordance with the theoretical analysis of the sequences of information backflow that is reported in the present Section.

\section{Summary and conclusions}\label{6}

In local dephasing channels, non-Markovian evolution and the process of recoherence of the qubit coincide with the appearance of information backflow \cite{MPRAr2013,MPRA2014,QbtMPRA2014,nnMarkovNeg2LZPRA2011,FPRA2013}. Several approaches have been developed to control the flow of quantum information. External interactions that are time-dependent, like dynamical decoupling pulses, allow the manipulation of the information flow \cite{MNJP2015,DDV1,DDV2}. The appearance of information backflow is strictly related to the structure of the reservoir of frequency modes. Consequently, engineering the external environment is certainly an effective way of controlling the information flow, the loss of coherence and the appearance of non-Markovian evolution. Ohmic-like spectra can be engineered in cold environments \cite{SDeng1,SDengBECM2011}. For ohmic-like SDs, information backflow appears uniquely if the ohmicity parameter exceeds the value $2$ \cite{MPRAr2013,MPRA2014}. 

In the present study we have considered structured reservoirs with low-frequency gaps in the continuous distributions of frequency modes \cite{KJMO1994,GPRA1997,GSDGPRA2012}. The spectral gap generates, along with non-Markovian evoution, a sequence of infinite time interval over which information backflow and recoherence appear. 
The SDs under study include the inverse-power-law divergences that characterize photonic band gap materials. Perturbations of the power-law profiles of the SDs have been considered by including arbitrary powers of logarithmic forms as possible additional factors \cite{GPRA2017,GXiv2016}. The time intervals over which information flows back in the open system are generally irregular. Still, under certain conditions that involve the SD, upper bounds for the starting and ending times of each episode of information backflow are determined by the cut-off frequency $\omega_g$ of the spectral gap. 
Each episode of information backflow lasts for an interval that is no longer than $\pi/\omega_g$. Over long times, the episodes of information backflow become regular. The corresponding time intervals are determined by the structure of the SD near the cut-off frequency $\omega_g$. Especially, the length of the long-time intervals tends to $\pi/\omega_g$ as supremum value. In conclusion, engineering a spectral gap and the low-frequency structure in the reservoir of frequency modes, produces in local dephasing channels, along with non-Markovian evolution, a sequence of regular and controllable long-time intervals over which information backflow and recoherence appear.

\appendix
\section{Details}\label{A}

The dephasing rate $\gamma_0(t)$ is defined by Eq. (\ref{gamma0g}). The change of variable $\omega=\omega_g + \omega^{\prime}$ leads to the forms (\ref{gammaSC}) and (\ref{gammaS}). The choice of the constraint (\ref{conSDphisp}) and of the forms (\ref{j1}) and (\ref{j2}) of the SDs is based on the properties of the sine transform \cite{TFI}. The analysis of the functions $\varphi_c(t)$ and $\varphi_s(t)$ is performed by considering Eq. (\ref{gamma0g}), and the relations $\varphi_c(t)=\omega^2_s f_c\left(\tau\right)/\omega_g$ and 
$\varphi_s(t)=\omega^2_s f_s\left(\tau\right)/\omega_g$. The dimensionless functions $f_c\left(\tau\right)$ and $f_s\left(\tau\right)$ are given by
\begin{eqnarray}
&&\hspace{-1em}f_c\left(\tau\right) = \int_0^{\infty}\Lambda\left(\nu\right) \cos\left(\nu \tau\right) d\nu, \label{varphic1ad} \\
&&\hspace{-1em}f_s\left(\tau\right) = \int_0^{\infty}\Lambda\left(\nu\right)\, \sin\left(\nu \tau\right) d\nu,
\label{varphis1ad}
\end{eqnarray}
where $\tau=\omega_s t$. The dimensionless function $\Lambda\left(\nu\right)$ is defined in terms of the auxiliary function $\Omega\left(\nu\right)$ in Sec. \ref{SD1}.
For the first class of SDs, introduced in Sec. \ref{SD1}, the function $\Lambda\left(\nu\right)$, defined by Eq. (\ref{Lambda}), behaves for $\nu\to 0^+$ as
\begin{eqnarray}
\hspace{-2em}
\Lambda\left(\nu\right)\sim\sum_{r=0}^{\infty}
\sum_{k=0}^{n^{\prime}_r}c^{\prime}_{r,k} \nu^{\alpha^{\prime}_r}\left(- \ln \nu\right)^k. \label{Ls01}
\end{eqnarray}
The powers $\alpha^{\prime}_r$ and $n^{\prime}_r$ and the coefficients $c^{\prime}_{r,k}$ are defined via the following product of series,
\begin{eqnarray}
\sum_{r=0}^{\infty}
\sum_{k=0}^{n^{\prime}_r}c^{\prime}_{r,k} \nu^{\alpha^{\prime}_r}\left(- \ln \nu\right)^k=
\sum_{l=0}^{\infty}\left(-\,\frac{\nu}{\nu_0}\right)^l\sum_{j=0}^{\infty}\sum_{k=0}^{n_j}c_{j,k} \nu^{\alpha_j}\left(- \ln \nu\right)^k. \label{seriesprod1}
\end{eqnarray}
Due to the properties of the powers $\alpha_j$, the following relations hold, $\alpha_{0}=\alpha_{0}^{\prime}$, $\alpha_{r+1}^{\prime}>\alpha_{r}^{\prime}$ for every $r=0,1,2,\ldots$, $\alpha_{r}^{\prime}\uparrow \infty$ as $r\to\infty$, $n_0=n^{\prime}_0$ and $c_{0,k}=c^{\prime}_{0,k}$ for every $k=0,\ldots,n_0$. 
For the second class of SDs, introduced in Sec. \ref{SD2}, the function $\Lambda\left(\nu\right)$ behaves for $\nu\to 0^+$ as
\begin{eqnarray}
\hspace{-2em}
\Lambda\left(\nu\right)\sim\sum_{r=0}^{\infty}
w^{\prime}_{r} \nu^{\alpha^{\prime}_r}\left(- \ln \nu\right)^{\beta^{\prime}_r}. \label{Ls02}
\end{eqnarray}
The powers $\alpha^{\prime}_r$ and $\beta^{\prime}_r$ and the coefficients $c^{\prime}_{r,k}$ are defined via the following product of series,
\begin{eqnarray}
\sum_{r=0}^{\infty}
w^{\prime}_{r} \nu^{\alpha^{\prime}_r}\left(- \ln \nu\right)^{\beta^{\prime}_r}
=\sum_{l=0}^{\infty}\left(-\,\frac{\nu}{\nu_0}\right)^l
\sum_{j=0}^{\infty}
w_j \nu^{\alpha_j}\left(- \ln \nu\right)^{\beta_j}. \label{seriesprod2} 
\end{eqnarray}
Due to the properties of the powers $\alpha_j$, the following relations hold, $\alpha_{0}=\alpha_{0}^{\prime}$, $\alpha_{r+1}^{\prime}>\alpha_{r}^{\prime}$ for every $r=0,1,2,\ldots$, $\alpha_{r}^{\prime}\uparrow \infty$ as $r\to\infty$, $\beta_0=\beta^{\prime}_r$ and $w_0=w^{\prime}_0$. 

The short- and long-time behavior of the functions $f_c(t)$ and $f_s(t)$ have been analyzed in Ref. \cite{GXiv2016} in terms of the asymptotic behaviors of the function $\Lambda\left(\nu\right)$ for $\nu\to 0^+$ and $\nu \to \infty$. Various constraints are required on the asymptotic behavior of the Mellin transform of the function $\Lambda\left(\nu\right)$ and certain integrals and derivatives that involve the function $\Lambda\left(\nu\right)$. See Ref. \cite{GXiv2016} for details. The analysis performed in such reference holds also for $0\geq\alpha_0>-1$. In case $\alpha_0=0$, the logarithmic power $n_0$, for the first class of SDs, or $\beta_0$, for the second class of SDs, must vanish, $n_0=\beta_0=0$, due to the constraint of summability. Following the results of Ref. \cite{GXiv2016}, we find that if the SDs belong to the first class, Sec. \ref{SD1}, and $\chi_0>1$, or to the second class, Sec. \ref{SD1}, and $\chi_0>3$, the functions $\varphi_c(t)$ and $\varphi_s(t)$ evolve algebraically for $t \ll 1/\omega_s$,
\begin{eqnarray}
&&\varphi_c(t)\sim l_{c,0}-l_{c,2} t^2,\hspace{1em}\varphi_s(t)\sim l_{s,1} t, \label{phscshort}
\end{eqnarray}
where
\begin{eqnarray}
&&l_{c,0}= \int_{\omega_g}^{\infty}\frac{J\left(\omega\right)}{\omega}\, d\omega, \hspace{1em} l_{s,1}= \int_{\omega_g}^{\infty}\frac{J\left(\omega\right)}{\omega}\,\left(\omega-\omega_g\right) d\omega, \nonumber \\
&&l_{c,2}= \frac{1}{2}\int_{\omega_g}^{\infty}\frac{J\left(\omega\right)}{\omega}\,\left(\omega-\omega_g\right)^2 d\omega.  \nonumber 
\end{eqnarray}
In this way, Eq. (\ref{gamma0short0}) is obtained, where $l_0=\omega_gl_{c,0}+l_{s,1}$. Following Ref. \cite{GXiv2016}, the long-time behaviors of the functions $\varphi_c(t)$ and $\varphi_s(t)$ are found in terms of the powers $\alpha^{\prime}_r$ and $n^{\prime}_r$, for the first class of SDs, or $\beta^{\prime}_r$, for the second class of SDs, which are defined via Eqs. (\ref{Ls01})-(\ref{seriesprod2}).

If $\varphi^2_s(t)+\varphi^2_c(t)>0$ for every $t\geq 0$, the expression (\ref{gammaS}) is obtained from Eq. (\ref{gammaSC}) via straightforward trigonometric relations \cite{GradRyz} between the functions $\operatorname{arccot}(x)$ and $\sin(x)$ or $\cos(x)$. The limit $\phi \left(\infty\right)$ of the time-dependent angle $\phi(t)$ for $t \to \infty$ is obtained from Eqs. (\ref{phi0})-(\ref{phi3}) and from the long-time behavior of the functions $\varphi_c(t)$ and $\varphi_s(t)$, given by Eqs. (\ref{varphic1})-(\ref{C2tInf0SD6}). In this way, Eqs. (\ref{phiinf0})-(\ref{phiinf5}) are obtained. The sign of the dephasing rate is analyzed from Eq. (\ref{gammaS}). The dephasing rate is negative over times $t$ such that
\begin{equation}
\pi \left(1+2n\right)<\omega_g t+\phi(t)<2\pi(1+n),
\label{negang1}
\end{equation}
for every $n=0,1,2,\ldots$.
Consequently, the constraint (\ref{Cepsilon0}) and Eqs. (\ref{t1n})-(\ref{dt2n}) estimate the long-time intervals over which the dephasing rate is negative. This concludes the demonstration of the present results.

\end{document}